\documentclass[aps, prb, preprint, superscriptaddress]{revtex4}
\usepackage{graphicx}
\usepackage{dcolumn}
\usepackage{bm}
\usepackage{natbib}
\usepackage{amsmath}
\usepackage{amssymb}
\usepackage{lipsum}
\usepackage{textcomp}
\usepackage{color}
\usepackage{siunitx}

\newcommand*{\vect}[1]{\mathbf{#1}}

\begin{document}

\title{Magnetic equivalent of electric superradiance: radiative damping in yttrium-iron-garnet films}

\author{L. Weymann}
\author{A. Shuvaev}
\author{A. Pimenov}
\affiliation{Institute of Solid State Physics, TU Wien, 1040 Vienna, Austria}
\author{A. A. Mukhin}
\affiliation{A. M. Prokhorov General Physics Institute of Russian Academy of Sciences, 119991 Moscow, Russia}
\author{D. Szaller}
\affiliation{Institute of Solid State Physics, TU Wien, 1040 Vienna, Austria}


\date{\today}

\maketitle
\textbf{A dense system of independent oscillators, connected only by their interaction with the same cavity excitation mode, will radiate coherently, which effect is termed superradiance. In several cases, especially if the density of oscillators is high, the superradiance may dominate the intrinsic relaxation processes. This limit can be achieved, e.g., with cyclotron resonance in two-dimensional electron gases. In those experiments, the cyclotron resonance is coupled to the electric field of light, while the oscillator density can be easily controlled by varying the gate voltage. However, in the case of magnetic oscillators, to achieve the dominance of superradiance is more tricky, as material parameters limit the oscillator density, and the magnetic coupling to the light wave is rather small. Here we present quasi-optical magnetic resonance experiments on thin films of yttrium iron garnet. Due to the simplicity of experimental geometry, the intrinsic damping and the superradiance can be easily separated in the transmission spectra. We show that with increasing film thickness, the losses due to coherent radiation prevail the system's internal broadening.}

Since the current CMOS-based electronics is soon to reach its limitations~\cite{delalamo_nature_2011}, the design of fundamentally new ways to forward, process, and store information is of vital importance. One possible direction is offered by magnonics, where information is manifested in the magnetic state of matter and forwarded as an oscillating magnetic wave (magnon)\cite{chumak_natphys_2015}. The lifetime of such an excited state, usually in the microwave frequency range, is a crucial factor when designing potential applications. Besides their use in information technology, the research of magnonic systems with long-lifetime excitations is also motivated by their role as model systems of fundamental physical effects, such as Bose-Einstein condensation and other macroscopic quantum transport phenomena~\cite{demokritov_nature_2006,bozhko_natphys_2016}.

Lifetime measurement of various excited states is an essential tool for analyzing physical, chemical, and biological processes and making spectroscopic state assignments~\cite{demas_book}. The most direct way is to observe the transient optical signal following the excitation pulse, typically in luminescence, and by fitting the exponential decay, the lifetime of the corresponding excited state can be determined~\cite{demas_jce_1976}. Another approach involves the phase shift of the response as compared to the modulated excitation signal~\cite{spencer_jcp_1970}.

The third method utilizes Heisenberg's uncertainty principle, which connects the lifetime of a state to the uncertainty of its energy~\cite{gislason_pra_1985}, resulting in the natural line broadening of the spectroscopic absorption or emission signal. However, starting from the earliest spectroscopic experiments, the intrinsic natural broadening is dominated by other phenomena, such as the collision between particles and the Doppler-effect due to thermal motion in the atomic spectral lines of gases~\cite{voigt_kbaw_1912}. In case of microwave magnetic resonance measurements, where the magnetic sample is coupled to the GHz radiation typically by a coplanar waveguide, the two main contributions to the non-intrinsic linewidth result from eddy currents induced either in the conducting sample~\cite{pincus_pr_1960} or in the waveguide~\cite{wende_pss_1976}, which effects are termed eddy-current- and radiative damping, respectively~\cite{schoen_prb_2015}.

In several cases, when intrinsic damping is low and the density of oscillators is high, the damping due to coherent radiation starts to dominate, the effect known as superradiance~\cite{cong_josab_2016}. Superradiance denotes coherent emission of uncoupled emitters when interacting with the same mode of electromagnetic wave, predicted in the Dicke-modell~\cite{dicke_pr_1954} and observed both in gases inside of an optical cavity~\cite{skribanowitz_prl_1973,gross_prl_1979,kaluzny_prl_1983}, in metamaterials~\cite{sonnefraud_acsnano_2010,wenclawiak_apl_2017}, and in two-dimensional electron gases~\cite{zhang_prl_2014, laurent_prl_2015, zhang_np_2016, maag_np_2016}. The synchronous decay of the excited emitters takes the form of a superradiant emission pulse, which shows a characteristic scaling with the density of the emitters. Namely, both the pulse amplitude and the corresponding damping rate (inverse lifetime) grow linearly with the emitter density. Depending on the coherently prepared/spontaneous origin of the initial coherent excited state of the emitters, the superfluorescence/superradiance terminologies are established in the literature, respectively.\cite{cong_josab_2016} Solid-state realizations, such as molecular centers in a crystal\cite{florian_pra_1984}, semiconductor quantum dots\cite{scheibner_natphys_2007} and quantum wells\cite{noe_natphys_2012}, high-mobility two-dimensional quantum gases\cite{zhang_prl_2014, laurent_prl_2015, zhang_np_2016, maag_np_2016} and nitrogen-vacancy centers in diamond\cite{bradac_natcomm_2017,angerer_natphys_2018}, provide experimentally more accessible ways to study superradiance under controlled temperature or external fields.  

In all examples of superradiance listed above the light-matter interaction takes place in the electronic channel, while the superradiance of magnetic resonances seems to be more challenging to realize. Nevertheless, superradiant response of nuclear spins has been found\cite{kiselev_mplb_1988} on ms timescale at very low, $T=0.3\textrm{ K}$ temperature and electron-spin superradiance of molecular magnets has also been proposed\cite{chudnovsky_prl_2002,tokman_prb_2008,stepanenko_scst_2016}, but has not been observed so far. In both cases, a passive resonant electric circuit around the sample is necessary to produce the oscillating magnetic field which builds up the coherence of the relaxation of individual spins.\cite{yukalov_lapl_2005,yukalov_ppn_2004} Recently, another magnetic alternative is proposed\cite{li_science_2018} where independent rare-earth magnetic moments are interacting with the spinwaves of the antiferromagnetically ordered iron spins in a magnetoelectric crystal. Here the expected superradiance of the rare-earth moments would appear as secondary-excited magnons of the iron system.

In this work, we present the optical transmission investigations of possible superradiance at the magnetic resonance in thin films of yttrium iron garnet (YIG). Based on Maxwell's equations, we separate the effects of the geometrical and magnetic parameters of the sample on the broadening of the resonance signal. Thus, the intrinsic damping due to the limited lifetime of the magnon excitation can be restored by following the frequency dependence of the absorption linewidth. Since the magnetic resonance is excited coherently, the radiative broadening, caused by the re-emission of a secondary electromagnetic wave, is also originating from a coherent process, showing similarities to the superradiance broadening effect observed in the same frequency-range cyclotron resonance of a three-dimensional topological insulator~\cite{gospodaric_prb_2019}. Unlike other proposals of electron-spin superradiance\cite{chudnovsky_prl_2002,yukalov_lapl_2005,yukalov_ppn_2004,tokman_prb_2008,stepanenko_scst_2016}, our method does not require a surrounding passive electronic resonator circuit, since the amplification of the radiated mode is produced by the sample itself, as the substrate gadolinium gallium garnet (GGG) layer plays the role of the resonance cavity.

\section*{Results}
\textbf{Magnon damping in thin magnetic films.} 
To address the effects of the intrinsic and radiative decay processes of magnons in thin magnetically ordered films, in the following a quasi-classical formalism is presented. The coherent magnetization dynamics is described by the Landau-Lifshitz-Gilbert model\cite{gurevich_book}, while the electrodynamical problem of optical transmission is treated within the thin sample approximation\cite{oksanen_jewa_1990, szaller_psr_2019} utilizing the boundary conditions resulting from Maxwell's equations. Thus, the resulted formulas for damping inherently assume a collective decay of the magnetic excitation. In the case of our sample, YIG, the choice of a coherent model is justified by the strong exchange coupling of the individual moments, which leads to their coherent motion in the long-wavelength limit. The classical formalism captures all features of superradiance\cite{eberly_ajp_1972}, only the assumption of an intrinsic damping, i.e. non-infinite lifetime of magnetic excitations, relays on quantum mechanics. However, an initial decay process is necessary to trigger the collective emission of the superradiant wave, while, on the other hand, the intrinsic decay should not dominate the radiation damping. Similarly to the case of cyclotron superradiance in semiconducting films~\cite{cong_josab_2016, gospodaric_prb_2019}, the radiative damping effect can only be observed in samples which are thin compared to the radiation wavelength. In the opposite case of the thick sample, the "superradiated" wave is re-absorbed, thus forming the propagation wave within the material.

The response of the magnetic moment $\vect{M}$ of YIG to an external magnetic field $\vect{H}$ is given by the Landau-Lifshitz-Gilbert equation of motion~\cite{gurevich_book}
\begin{equation}
\frac{\text{d}\vect{M}}{\text{d}t} = -\gamma \mu_0 \left( \vect{M} \times \vect{H} \right) + \frac{\alpha}{M_0}\left(\vect{M}\times\frac{d\vect{M}}{\text{d}t}\right) \, \text{,}
\label{eqMot}
\end{equation}
where $\gamma$ denotes the gyromagnetic ratio, $M_0$ the saturation magnetization, $\alpha$ the dimensionless Gilbert damping parameter, and $\mu_0$ the vacuum permeability.

In case of an incoming THz beam with angular frequency $\omega$ and within Faraday geometry (wave propagation is parallel to the magnetic field), the magnetic field in the film is given by $\vect{H}=\left( h_\text{x} e^{i\omega t}, h_\text{y}e^{i\omega t}, H_0^\text{F}\right)$. Here $h_\text{x,y}e^{i\omega t}$ is the oscillating (AC) field in the sample plane,  $H_0^\text{F}$ is the strength of the static (DC) magnetic field in the material given by $H_0^\text{F}=|H_0-M_0|$, with the applied external field $H_0$. Linearizing Eq.\,(\ref{eqMot}), we finally get the standard result for the
magnetic susceptibility for the circularly polarized light in the Faraday geometry:
\begin{equation}
\chi_\pm^\text{F} = \frac{m_\pm}{h_\pm} = \frac{\gamma M_0}{\omega_0^\text{F}\mp\omega+i \alpha \omega}\,\text{,}\label{chi_F}
\end{equation}
where $m_\pm, h_\pm$ are the circularly polarized AC magnetization and field, respectively, and $\omega_0^\text{F} = \gamma |H_0 - M_0|$
is the angular frequency of the ferromagnetic resonance.

Similarly, in the Voigt geometry (wave propagation perpendicular to the magnetic field) with linearly polarized light $\vect{H}=(H_0, h_\text{y} e^{i\omega t}, 0)$ the susceptibility is obtained as
\begin{equation} \label{chi_V}
\chi^\text{V}_y = \frac{(\omega_0^\text{V})^2 {M_0}/{H_0}} {\left(\omega_0^\text{V}\right)^2+\left(\alpha-i\right)^2\omega^2} \, ,
\end{equation}
with $\omega_\text{0}^\text{V} = \gamma \sqrt{H_0(H_0 + M_0)}$ resonant angular frequency.

To obtain the radiative contribution to the damping, we consider the transmission through a magnetic sample in the thin film approximation~\cite{oksanen_jewa_1990, szaller_psr_2019}. In this approximation, the boundary conditions are rewritten, taking the thin film as a part of the boundary.
For the Faraday geometry with circularly polarized light, we use the Maxwell equations
\begin{equation}\label{eq_maxwell}
\begin{aligned}
  \oint_{\partial A}\vect{E}\cdot \text{d}\vect{l}&=& -\iint_A\frac{\partial\vect{B}}{\partial t}\cdot d\vect{A}\\
	\oint_{\partial A}\vect{H}\cdot \text{d}\vect{l}&=& \iint_A\frac{\partial\vect{D}}{\partial t}\cdot d\vect{A}
\end{aligned}
\end{equation}
to obtain the fields on both sides of the film:
\begin{equation} \label{eq_boundary}
\begin{aligned}
 e^0_\pm-e_\pm^\text{r}+e_\pm^\text{br}-e^\text{t}_\pm &= i\omega d \mu_0\left(h_\pm+m_\pm\right)\\
 h^0_\pm+h^\text{r}_\pm+h_\pm^\text{br}-h^\text{t}_\pm &= i\omega d e_\pm\varepsilon\varepsilon_0 \, .
\end{aligned}
\end{equation}

On the left-hand side of Eq.\,(\ref{eq_boundary}), ($e$) and ($h$) are the AC electric and magnetic fields of the incident (0), reflected (r) and transmitted (t) wave. The reflected wave is partially back-reflected from the other surface of the substrate, as indicated by the $e_\pm^\text{br}=f(\omega) e_\pm^\text{r}$, $h_\pm^\text{br}=f(\omega) h_\pm^\text{r}$ terms, where $f(\omega)$ shows to the changes in the phase and amplitude due to the twice propagation trough the substrate and due to the reflection from the substrate-vacuum interface. In our experiments, considering the refractive index and absorption coefficient of the GGG substrate, this term causes a frequency-dependent modulation  of the transmission amplitude with $\lvert f(\omega)\rvert\sim 0.2$ relative amplitude due to the Fabry-P\'erot interference. On the right-hand side of Eq.\,(\ref{eq_boundary}), within the thin-sample approximation, the linear dependence of the AC electric and magnetic fields along the surface normal of the sample is assumed. Thus, $e_\pm$, $h_\pm$ and $m_\pm$ are the mean values of the corresponding fields and of the magnetization inside the film. Here $d$ denotes the film thickness, and $\varepsilon$, $\varepsilon_0$ stand for the electric permittivity of YIG and vacuum, respectively. To obtain Eq.\,(\ref{eq_boundary}), the integration path in the Maxwell equations must be taken across both sides of the sample. 

The AC electric and magnetic fields are connected via $e=Z_0 h/n_{1,2}$, where $Z_0=\sqrt{\mu_0/\varepsilon_0}$ is the impedance of free space, and $n_{1,2}$ is the refractive index of the dielectric media on both sides of the film. To further simplify the final expressions, for frequencies close to the ferromagnetic resonance, we assume $\chi_\pm^\text{F}\gg 1$ and neglect all smaller terms. After some simple algebra, the transmission coefficient is obtained as:
 \begin{equation} \label{eq_tran}
 t_{\pm}^\text{F}=\frac{h_\pm^\text{t}}{h^0_\pm}=\left(\frac{n+1-f(\omega)(n-1)}{2}+\frac{i\omega n d\mu_0\chi_\pm^\text{F}(1-f(\omega))}{2Z_0}\right)^{-1}\, , 
 \end{equation}
where $n$ stands for the refractive index of the substrate (GGG in our case).

In order to obtain relative transmission due to the magnetic resonance in YIG, Eq.\,(\ref{eq_tran}) should be normalized by the transmission through the pure substrate, $t|_{d=0}$ (see also Note\,[\onlinecite{note}]):
\begin{equation}\label{eq_reltran}
\frac{t_\pm^\text{F}}{t_\pm^\text{F}|_{d=0}}=
1-\frac{i\Gamma^\text{F}_\text{rad}}{\omega_0\mp\omega+i\alpha\omega+i\Gamma^\text{F}_\text{rad}} \, ,
\end{equation}
where we introduce the radiative damping parameter as
\begin{equation}
\Gamma^\text{F}_\text{rad}=\frac{n}{Z_0(n+1)}\omega d\mu_0\gamma M_0 \left( 1-\frac{2}{n+1}f(\omega)\right) = 2\pi\frac{ d}{\lambda}\gamma M_0\frac{n}{n+1}\left( 1-\frac{2}{n+1}f(\omega)\right) \text{.}\label{eq_gammaSR}
\end{equation}
Here $\lambda$ is the radiation wavelength in vacuum.

In the case of Voigt geometry, the normalized transmission for a linearly polarized wave with $h_{y}$ results in the same expression as in Eq.\,(\ref{eq_reltran}), where the superradiant damping parameter is replaced by
\begin{equation}\label{eq_gammaVoigt}
\Gamma_\text{rad}^\text{V}=\frac{\Gamma_\text{rad}^\text{F}}{2}\, .
\end{equation}

\begin{figure}
\centering{}\hbox{\hspace{-0cm}\includegraphics[width=1\textwidth]{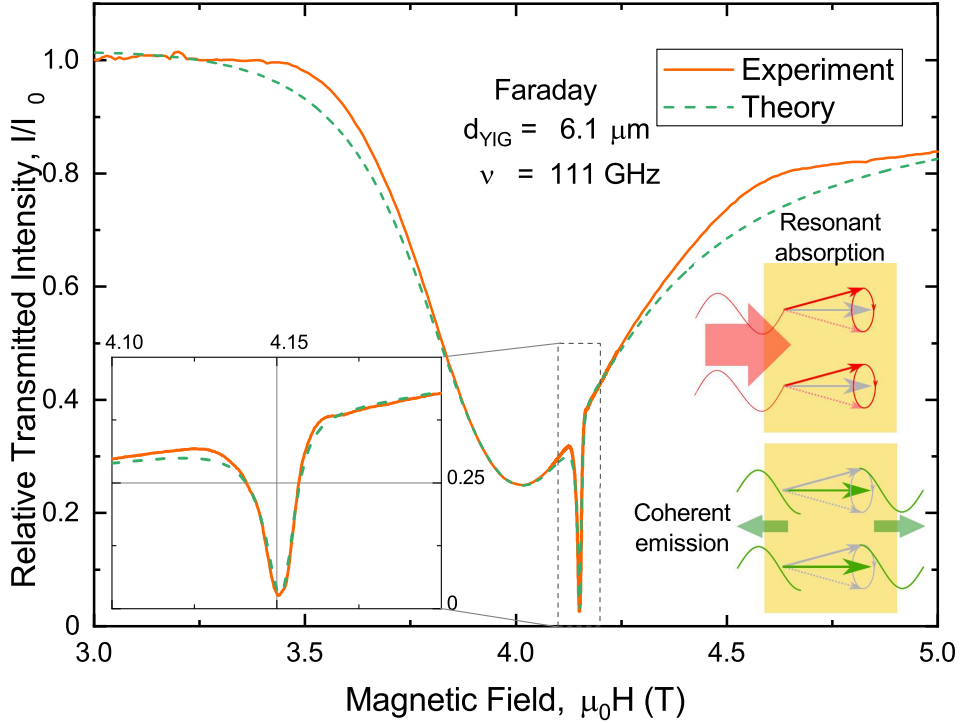}}
\caption{\label{fieldScan}
\textbf{Magnetic field dependence of the transmission of a YIG/GGG system at $\nu=111\textrm{ GHz}$.} The transmitted light intensity $I$ is shown relative to the GGG substrate's transmission baseline $I_0$ far from the paramagnetic resonance. The solid orange curve shows the experiment and the dashed green line the fit using the Fresnel equations\cite{shuvaev_sst_2012}. The broad minimum around a magnetic field of 4.0\,T corresponds to the paramagnetic resonance in GGG, while the sharp minimum at 4.15\,T is due to the ferrimagnetic resonance in YIG. The inset shows the expanded view of this resonance in YIG. The underlying physical processes, namely magnetic absorption and coherent emission, are depicted in the schematic cartoons. } \label{fig_tr}
\end{figure}

For comparison, in the case of electric superradiance in thin films the corresponding equation reads~\cite{gospodaric_prb_2019}
\begin{equation}
\Gamma_{\text{rad}} = n_\text{2D}e^2 Z_0/2m \, .
\end{equation}
Here $n_\text{2D}$ is the density, $e$ the charge, and $m$ the effective mass of two-dimensional electrons. In the electric case, to control the ratio between intrinsic and radiation damping, the electron density can be easily changed by variation of the gate voltage~\cite{gospodaric_prb_2019}. In the magnetic case the parameter responsible for the radiance intensity is the static magnetization $M_0$ that can be modified by tuning the temperature. An alternative way to control the superradiance is by changing the sample thickness $d$, as seen in Eq.\,(\ref{eq_gammaSR}).

Equation~(\ref{eq_reltran}) provides an expression for a resonant minimum in transmission with an effective width given by the sum of intrinsic$(\alpha\omega)$ and radiative$(\Gamma_\text{rad}^\text{F})$ parts. Since both the amplitude and the linewidth of the peak in Eq.\,(\ref{eq_reltran}) contains $\Gamma_\text{rad}^\text{F}$, the peak power of the secondary re-emitted wave grows with the square of the number of magnetic moments due to $\Gamma_\text{rad}^\text{F}\propto M_0 d$. This scaling is characteristic for coherent radiation, such as superradiance.

A quantum-mechanical approach of magnonic superradiance\cite{shrivastava_jpcssp_1976}, representing the interacting magnon-photon system by bosonic operators, results basically the same formula for the radiation damping as our classical method in Eq.\,(\ref{eq_gammaSR}) in the case of a small magnetic sample placed in a resonance cavity terminated at one end. Identifying the cavity of Ref.[\onlinecite{shrivastava_jpcssp_1976}] with the substrate GGG layer of our study allows a one-to-one comparison of the calculated radiation damping rates. The scaling with the sample parameters and the wavelength, $\Gamma_{\text{rad}}\propto n\frac{ d}{\lambda}M_0$, corresponds to our findings. However, the quantum-mechanical model\cite{shrivastava_jpcssp_1976} also proposes a linear increase of the radiation damping parameter with the $N$ density of photons in the sample, $\Gamma_{\text{rad}}\propto (1+2N)$, originating from the commutation relation of bosonic operators. Experimentally this quantum-mechanical effect would correspond to a dependence of $\Gamma_{\text{rad}}$ on the intensity of the incoming radiation for high-intensity probing beam. Given the limited power of our light sources we could not observe the intensity dependence of the radiation damping. Thus, since in the low-intensity, $N<<1$ regime the classical formalism provides an equivalent description of radiative damping, in the following we apply the classical formulas.

\begin{figure}
	\centering{}\hbox{\hspace{0cm}\includegraphics[width=1\textwidth]{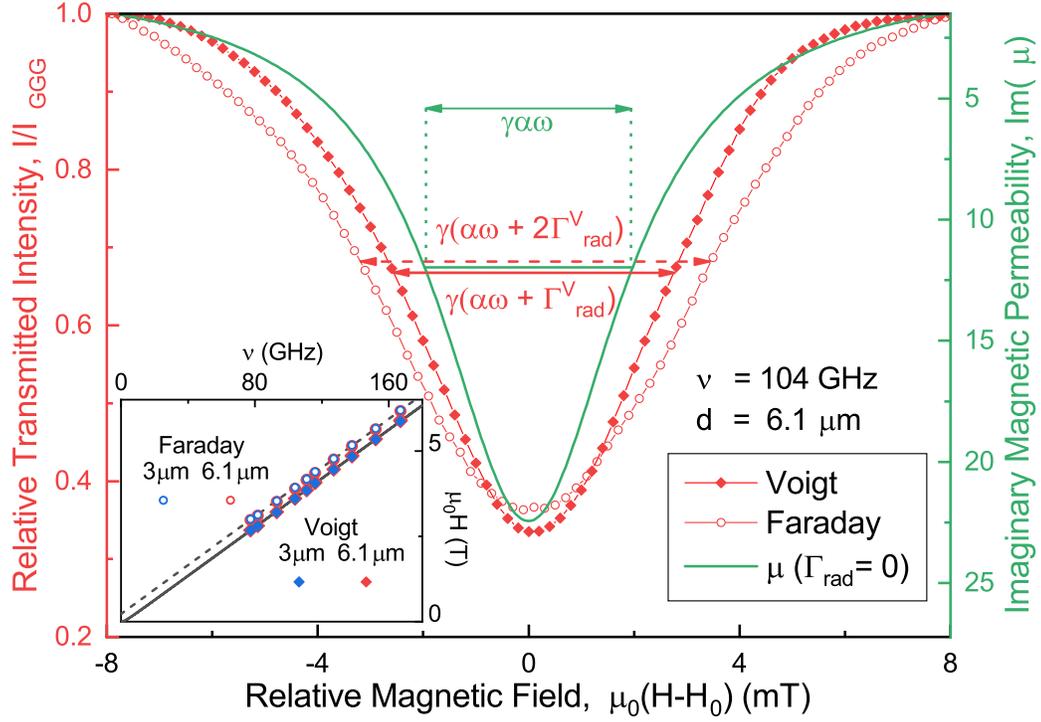}}
	\caption{\label{f_res}
	 \textbf{Ferrimagnetic resonance linewidth in YIG at $\nu=104\textrm{ GHz}$ frequency.} Resonance curves correspond to the intrinsic damping ($\mathrm{Im}(\mu)$, green) and to observed transmission in the Faraday (red open circles) and Voigt (red filled diamonds) geometries. The horizontal axis shows the magnetic field $H$ relative to the corresponding resonance field $H_0$. Red labels on the left vertical axis belong to the experimental curves and indicate the transmitted light intensity $I$ relative to the substrate's transmission baseline $I_{GGG}$, which can be measured without the YIG layer. The inverse scale on the green right axis corresponds to the imaginary part of the magnetic permeability. The frequency dependences of the ferrimagnetic resonance fields (open/filled symbols for Faraday/Voigt geometry experiments) in two YIG films of different thicknesses ($d=3\textrm{ }\mu\textrm{m}$ in blue and $d=6.1\textrm{ }\mu\textrm{m}$ in red) are presented in the inset, where dashed/solid lines show theoretical expectations of the resonance frequencies for Faraday/Voigt geometry. On the scale of the inset symbols corresponding to the two samples at a given frequency are completely superposed.}
\end{figure}

\textbf{Analysis of the different contributions to the absorption linewidth.} According to Eq.\,(\ref{eq_reltran}), compared to an intrinsic damping $\Gamma_\text{int}=\alpha \omega$, the resonance is additionally broadened by the radiative damping $\Gamma_{\mathrm{rad}}$. As illustrated in Fig.\,\ref{f_res}, the analysis of the linewidth and amplitude of the resonance according to Eq.\,(\ref{eq_reltran}) provides a way to directly obtain both damping parameters independently.

Fig.\,\ref{fig_tr}. shows the typical magnetic field dependent transmission of YIG films on GGG substrate. The sharp minimum at 4.14\,T that corresponds to the ferrimagnetic resonance of YIG is observed on top of the broad paramagnetic resonance of the GGG substrate. The two resonances were fitted simultaneously using Fresnel equations for the transmission of the layered system. The free parameters in such a fit are the saturation magnetization of the YIG film $M_0$, intrinsic damping $\alpha$, and the resonance field $H_\text{res}$. The refractive index of the GGG substrate, $n=3.43$, was obtained in a separate experiment~\cite{schneider_prl_2009}. The magnetic field dependence of the paramagnetic resonance frequency in GGG corresponds to a g-factor of $g_{GGG}= 2.26$.

The ferrimagnetic resonance field $H_\text{res}$ of YIG shows an approximately linear dependence on the angular frequency $\omega$ (see inset of Fig.\,\ref{f_res}), which is expected in high magnetic fields $H_0 \gg M_0$. From the fits to these data we estimate the value for the saturation magnetization $\mu_0 M_0(T=200\mathrm{ K}) = 0.2 \pm 0.03 \,$T, in reasonable agreement with  the literature values~\cite{anderson_pr_1964} and with the static data, $\mu_0 M_0^\text{VSM}= 0.198$\,T, measured on the same sample using vibrating sample magnetometry.

Figure \ref{f_res}. demonstrates that the ferrimagnetic resonance linewidth of YIG in the transmission is indeed substantially higher than the actual resonance linewidth in the magnetic permeability, $\mu(H)$. Moreover, the radiative correction in Faraday geometry is twice as large as in Voigt configuration, in agreement with Eq.\,(\ref{eq_gammaVoigt}). The transmission spectra allow us to extract the relevant electrodynamic parameters using different approaches. In a first approach it can be done using the \emph{exact} Fresnel expressions~\cite{shuvaev_sst_2012} including the magnetic permeability given by Eqs.\,(\ref{chi_F},\ref{chi_V}). Here the radiative damping is not a free parameter~\cite{gospodaric_prb_2019} but is obtained via Eq.\,(\ref{eq_gammaSR}). The second way to compare intrinsic and radiative damping is to use the simplified Eq.\,(\ref{eq_reltran}). In this case, the width of the resonance curve directly gives the total damping parameter $\Gamma_\text{int}+\Gamma_\text{rad}$, and the intrinsic damping is extracted from the amplitude of the resonance curves.

\begin{figure}
	\centering{}\hbox{\hspace{0cm}\includegraphics[width=0.5\textwidth]{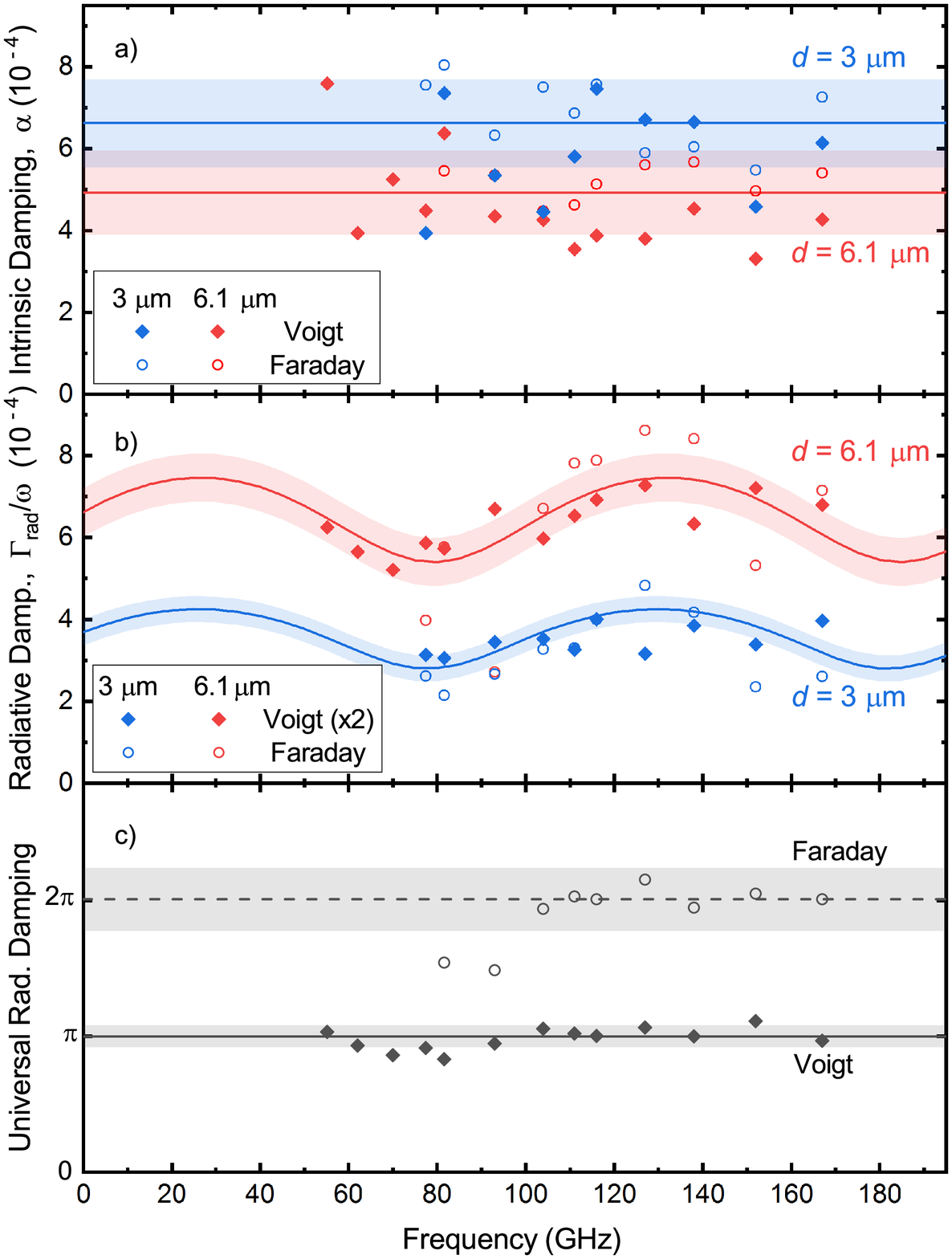}}
	\caption{\label{f_damp_freq} \textbf{Frequency dependence of the dimensionless intrinsic ($\alpha$, a) and radiative damping ($\Gamma_{rad}/\omega$, b,c) parameters.} The values are obtained from fits of the transmission  according to Eq.~(\ref{eq_reltran}). Damping rates of YIG samples with different thicknesses are indicated by color (6.1$\mu$m - red, 3$\mu$m - blue), while full/empty symbols correspond to Voigt/Faraday geometries, respectively. Radiative damping of the Voigt experiments in b) is scaled up by a factor of 2, according to Eq.\,(\ref{eq_gammaVoigt}), to allow a numeric comparison with the Faraday-geometry measurements. Mean values of the intrinsic damping parameters belonging to the two samples are indicated by constant lines, while standard deviations of the data sets are shown by bands of the corresponding colors (a). Frequency-dependent oscillations of the radiative damping parameters belonging to the two samples are indicated by fits of $\Gamma_{rad}/\omega\propto\left(1-\frac{n}{n+1}f(\omega)\right)$, while root mean squares of the deviations of the data points from the fits are shown by bands of the corresponding colors (b). Universal radiative damping rate, $\Gamma_{rad}/\left(\frac{ d}{\lambda}\gamma M_0\frac{n}{n+1}\left(1-\frac{n}{n+1}f(\omega)\right)\right)$, c, agrees with $\pi$ and $2\pi$ for Voigt (median shown by solid black line) and Faraday (median indicated by dashed black line) geometry measurements, respectively. Here symbols are obtained by averaging corresponding measurements on the two samples at the given frequency and gray bands show the standard deviation of the datasets.}
\end{figure}

The values of intrinsic damping obtained from the present experiments are shown in Fig.\,\ref{f_damp_freq}(a).
As expected, the $\alpha$ Gilbert damping parameters of our samples are approximately frequency-independent and agree reasonably well with each other and with former reports~\cite{chen_ieeetm_1989, kelly_apl_2013, hahn_apl_2014, liu_jap_2014, onbasli_aplmat_2014, haidar_prb_2016, jermain_prb_2017, fanchiang_natcomm_2018, boventer_prb_2018, pfirrmann_prr_2019, kosen_aplmat_2019}.

Figure \ref{f_damp_freq}(b) shows the frequency dependence of the dimensionless radiative damping parameter, $\Gamma_\text{rad}/\omega$, at the magnetic resonance field.
As expected from Eq.\,(\ref{eq_gammaSR}), $\Gamma_\text{rad}$ is proportional both to the
frequency and to the sample thickness. Moreover, the frequency-dependent oscillation of $\Gamma_\text{rad}/\omega$ corresponds to the expected behavior due to the Fabry-P\'erot interference of the direct light beam and the beam reflected from the substrate-vacuum surface. To numerically validate our model in Eqs.\,(\ref{eq_gammaSR},\ref{eq_gammaVoigt}), we compared the radiative damping rates, rid of the experimental and material parameters, to the universal values of $\pi$ and $2\pi$ for Voigt and Faraday geometries, respectively. As presented in Fig.\,\ref{f_damp_freq}(c), despite of the scattering of the individual data points, the median value of the universal radiative damping $\Gamma_{rad}/\left(\frac{ d}{\lambda}\gamma M_0\frac{n}{n+1}\left(1-\frac{n}{n+1}f(\omega)\right)\right)$ agrees reasonably well with the model. Comparing the values of the intrinsic (Fig.\,\ref{f_damp_freq}(a)) and radiative (Fig.\,\ref{f_damp_freq}(b)) damping parameters, we see that in the thinner film with $d=3\, \mu$m, the intrinsic damping dominates, while in the thicker film both contributions are of comparable values: $\alpha  \sim \Gamma_\text{rad}(d=6.1\,\mu \mathrm{m})/\omega $. Thus, in thicker films the observation of a relaxation in the form of a superradiant pulse might be possible.

\begin{figure}
	\centering{}\hbox{\hspace{0cm}\includegraphics[width=1\textwidth]{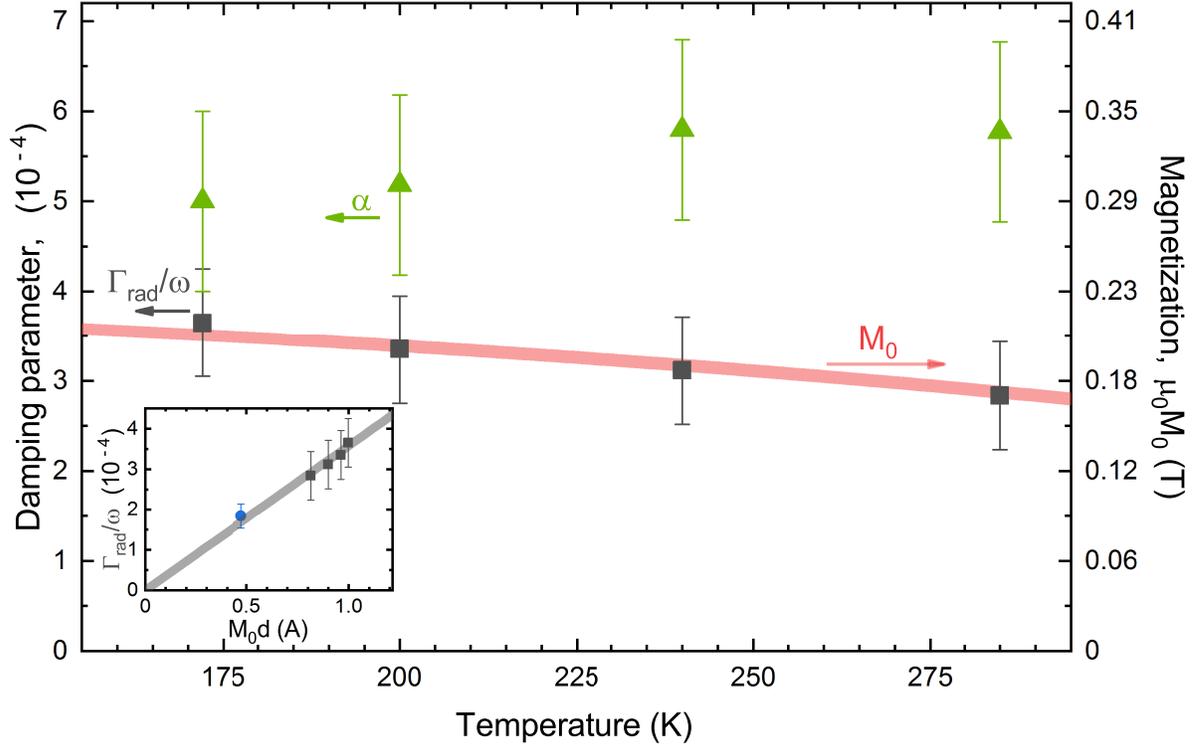}}
	\caption{\textbf{Temperature dependence of the damping parameters.} The left scale corresponds to the dimensionless radiative damping $\Gamma_{\mathrm{rad}}/\omega$ (black squares) and intrinsic damping $\alpha$ (green triangles) parameters in the temperature range $T=170-290\textrm{ K}$, as observed on the 6.1\,{\textmu}m thick YIG sample. Error bars represent the standard deviation of the measured values at various frequencies in the $\nu=40-190\textrm{ GHz}$ range. For comparison, red line shows the temperature dependence of the $M_0$ saturation magnetization of YIG, reproduced from Ref.\,[\onlinecite{maierflaig_prb_2017}], on the right scale. In the inset the radiative damping  versus the magnetic moment per unit area, $M_0d$, is presented. Black squares correspond to the 6.1\,{\textmu}m thick YIG sample, while blue circle to the 3\,{\textmu}m thick YIG film. Grey line shows a linear fit of the measured data.}\label{f_temp}
\end{figure}

The collective nature of the radiative relaxation is visible in the interference pattern of the frequency dependence of $\Gamma_\text{rad}/\omega$ in Fig.\,\ref{f_damp_freq}(b). Since the refractive indices of YIG\cite{sirdeshmukh_bms_1998} and that of the GGG~\cite{schneider_prl_2009} substrate agree within 5\%, the fields treated in the general description of Eq.\,(\ref{eq_boundary}) as reflected from the YIG-GGG interface in the reality correspond to the secondary radiation field of the YIG layer. Thus, the interference pattern in Fig.\,\ref{f_damp_freq}(b) produced by the secondary radiated field of YIG reflected back from the GGG-vacuum interface is a clear sign of the coherent secondary radiation. When comparing the frequency dependence of the intrinsic (Fig.\,\ref{f_damp_freq}(a)) and radiative (Fig.\,\ref{f_damp_freq}(b)) damping rates, the oscillations only appear in the latter case, corresponding to the model of coherent emitters in Eq.\,(\ref{eq_gammaSR}).    

Alternatively, the collective nature of the secondary radiation can be studied on the dependence of $\Gamma_\text{rad}$ on the number of emitters. Namely, for coherent emission the radiative damping parameter grows linearly with the magnetic moment per unit area of the quasi two-dimesional structure, $\Gamma_\text{rad}\propto M_0 d$. In Fig.\,\ref{f_damp_freq}(b) the proportionality of $\Gamma_\text{rad}\propto d$ is observed, while the temperature dependence of $\Gamma_\text{rad}$ is presented in Fig.\,\ref{f_temp}. With increasing temperature, the dimensionless radiation damping $\Gamma_\text{rad}/\omega$ is decreasing, while the intrinsic relaxation $\alpha$ increases. As both the temperature dependence of $M_0$ in Fig.\,\ref{f_temp} and the $M_0d$ dependence of $\Gamma_\text{rad}/\omega$ in the inset of Fig.\,\ref{f_temp} indicate, the $\Gamma_\text{rad}\propto M_0d$ relation holds in the experiments, indicating a collective radiation damping process. 

\section*{Discussion}

We have presented a method to separate the intrinsic and radiative contributions to magnetic thin films' resonance linewidth in optical absorption experiments. Compared to previous microwave studies\cite{kelly_apl_2013, hahn_apl_2014, liu_jap_2014, jermain_prb_2017}, the higher frequency $(\nu\sim 100\textrm{ GHz})$ of our optical approach provides a more precise way to determine the intrinsic Gilbert damping parameter $\alpha$. Moreover, the free-standing sample in transmission experiments offers a universal method to eliminate the damping effect of the measurement setup, such as the induced currents in the waveguide or resonator cavity. 

In the case of thin samples, the re-emission of the electromagnetic radiation by the individual magnetic moments occurs in a coherent process, resulting in a quadratic scaling of the emitted power with the magnetization of the material, and with the volume of the sample. The coherent nature of the radiative line broadening mechanism allows its identification as the magnetic equivalent of superradiance, opening a fundamentally new way to study this collective phenomenon in the dynamics of magnetic systems. The characteristic frequency of magnetic excitations is by several orders of magnitude lower than that of the other extensively investigated quantum-optical systems\cite{baumann_nature_2010}, granting the possibility of time-resolved detection of the magnetic superradiant dynamics.  

\section*{Methods} 

\textbf{Sample preparation.} In this work, we investigated two different YIG/GGG systems (Y$_3$Fe$_5$O$_{12}$ thin film on Gd$_3$Ga$_5$O$_{12}$ substrate) grown by liquid epitaxy. The first sample is a 3\,$\mu$m thick YIG film on 537\,$\mu$m GGG. The second sample consists of two 6.1\,$\mu$m YIG-films on both sides of a 471\,$\mu$m thick GGG substrate.

Since the thickness of the thin films is crucial for the interpretation of the optical experiments, for quality control, static magnetization measurements were performed in a vibrating sample magnetometer and in the temperature range of 5\,-\,300\,K. The thickness of the YIG layer in the sample was accurately determined from the magnitude of the observed magnetization step at a small magnetic field, corresponding to the switching of the ferrimagnetic magnetization of the YIG film.

\textbf{Sub-THz spectroscopy.} Quasi-optical experiments were performed in a sub-THz Mach-Zehnder interferometer~\cite{volkov_infrared_1985,kuzmenko_prl_2018}, equipped with a 7\,T magnet. Except of the temperature dependent study in Fig.\,\ref{f_temp}, during the other measurements the sample temperature was kept at 200\,K. We distinguish between the Faraday and the Voigt geometry, where the magnetic field was applied parallel and perpendicular to the direction of the incident light beam, respectively. The monochromatic radiation was generated by a set of backward-wave oscillators covering the frequency range of 40\,GHz\,-\,1\,THz. Metal wire-grids were used to achieve a linear polarization of light in Voigt geometry, while a combination of an additional wire grid and a mirror (producing a $\pi/2$ phase shift between two linear polarizations) was used to obtain circularly polarized light for the Faraday geometry.

\begin{figure}
	\centering{}\hbox{\hspace{0cm}\includegraphics[width=1\textwidth]{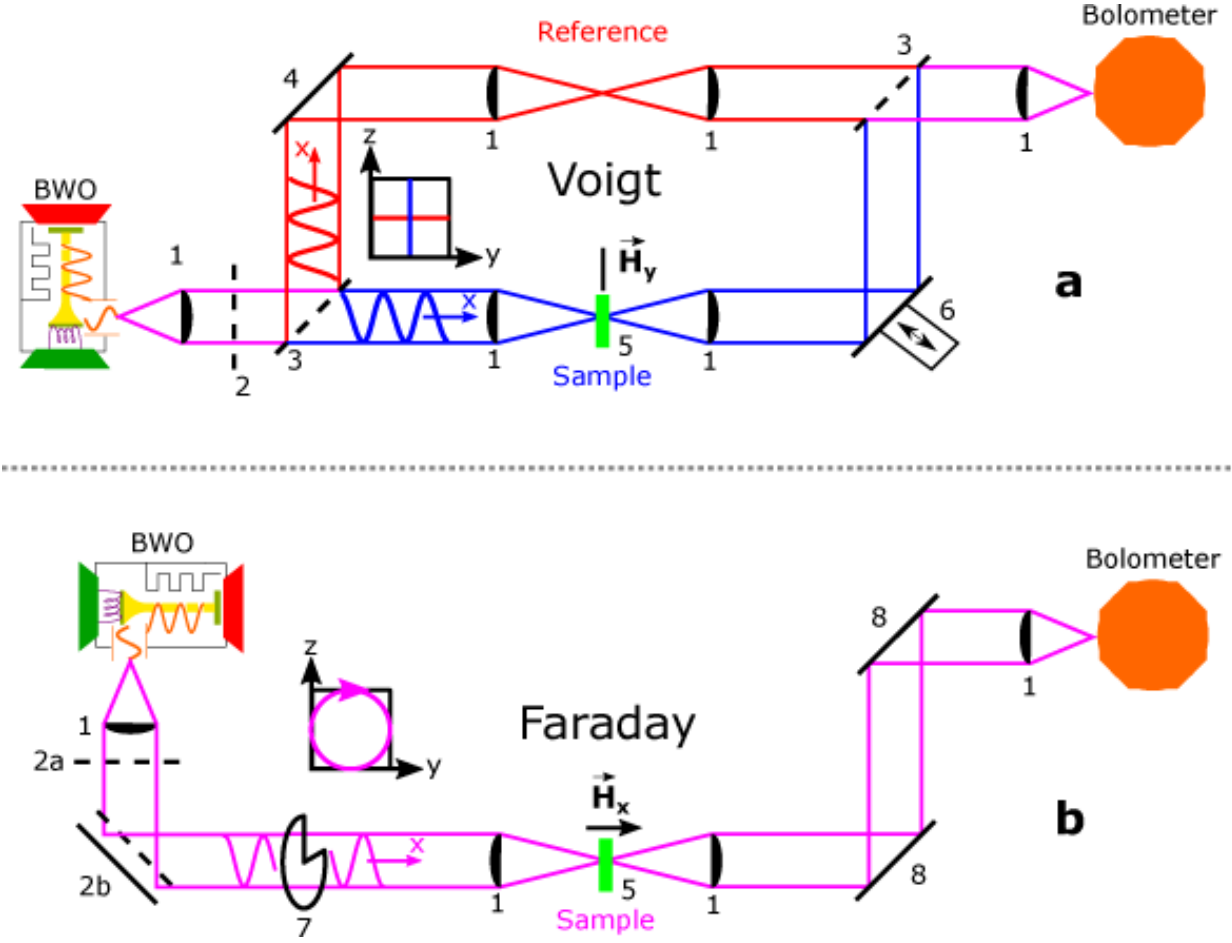}}
	\caption{\textbf{Experimental setup.} Mach-Zehnder interferometer using linearly polarized light beam in Voigt(a), and transmission measurement setup with circularly polarized beam in Faraday(b) geometry. The signal is detected by $T=4\textrm{ K}$ bolometer, BWO stands for backward-wave oscillator light source and optical elements are marked by numbers (1 PTFE focusing lens, 2, 2a linear polarizer, 2b circular polarizer, 3 beam-splitting polarizer, 4 standing mirror, 5 sample in magnetic field, 7 circularly polarized light, 8. standing mirror). Insets show the polarization state of the incoming light (linear for Voigt and circular for Faraday geometry).}\label{f_setup}
\end{figure}

\textbf{Data analysis.} The transmission spectra have been analyzed using the Fresnel optical equations for a multilayer system~\cite{shuvaev_sst_2012}, assuming a Lorentzian line-shape for the ferri/paramagnetic resonance in the magnetic permeability of the YIG film and the GGG substrate, respectively. As discussed above, to visualize the effects of magnetic superradiance, simplified expressions for the transmission coefficient are more appropriate. This procedure allows a direct estimate of the relevant material parameters from the spectra.

\section*{Acknowledgments}

The authors thank S. O. Demokritov for the YIG sample and M. Kramer for her assistance in fitting the spectra. This work was supported by the Austrian Science Funds (W 1243, I 2816-N27, P 27098-N27, TAI 334-N).

\bibliography{literature}

\begin{thebibliography}{61}
\bibitem[{not()}]{note}
\bibinfo{note}{In case of two films on both sides of the substrate the
  transmission in the same approximation is given by $t(2d)=[t(d)]^2$}.
\end{thebibliography}

\end{document}